\documentstyle[doublespacing,epsfig]{jaa}

\begin{document}

\title [Wide HI absorption toward Sgr A*]
       {On the Origin of the Wide HI Absorption Line Toward Sgr A*}

\author[Dwarakanath, Goss, Zhao \& Lang]
       {K. S. Dwarakanath,$^1$\thanks{e-mail:dwaraka@rri.res.in}
	W. M. Goss,$^2$\thanks{e-mail:mgoss@aoc.nrao.edu}
	J. H. Zhao$^3$\thanks{email:jzhao@cfa.harvard.edu}
	\& C. C. Lang$^4$\thanks{email:cornelia-lang@uiowa.edu}\\
       $^1$Raman Research Institute, Sadashivanagar, Bangalore 560 080 \\
       $^2$National Radio Astronomy Observatory, PO BOX O, Socorro, NM 87801\\
       $^3$Harvard-Smithsonian Center for Astrophysics, Cambridge, MA 02138 \\
       $^4$Department of Physics and Astronomy, University of Iowa, Iowa City, IA 52242}

\maketitle

\begin{abstract}

We have imaged a region of $\sim$ 5$^{'}$ extent surrounding Sgr A* 
in the HI 21 cm-line absorption using the Very Large Array. 
A Gaussian decomposition of the optical depth
spectra at positions within of $\sim$ 2$^{'}$ ($\sim$ 5 pc at 8.5 kpc) 
of Sgr A* detects a wide line underlying the many
narrow absorption lines. The wide line has a mean peak optical depth of
0.32 $\pm$ 0.12 centered at a mean velocity of V$_{lsr}$ = --4 $\pm$ 15 km s$^{-1}$. The
mean full width at half maximum is 119 $\pm$ 42 km s$^{-1}$. 
Such a wide line is absent in the spectra at positions beyond $\sim$ 2$'$
from Sgr A*. The position-velocity diagrams in optical depth reveal that the
wide line originates in various components of the circumnuclear disk 
(radius $\sim$ 1.3$'$) surrounding Sgr A*. These components
contribute to the optical depth of the wide line in different velocity ranges.  
The position-velocity diagrams do not reveal any diffuse feature 
which could be attributed
to a large number of HI clouds along the line of sight to Sgr A*.
Consequently, the wide line has no implications either to a global population
of shocked HI clouds in the Galaxy or to the energetics of the interstellar medium
as was earlier thought. 

\end{abstract}


\section{INTRODUCTION}

The Galaxy harbors a compact radio source Sgr A* at its center. The
intrinsic size of Sgr A* is constrained by the most recent multi frequency
VLBA observations to be 24$\pm$2 Schwarzschild radius ($\sim$ 2 AU for
a 4$\times$ 10$^{6}$ M$_\odot$ black hole) at 43 GHz and scales
as $\lambda ^{1.6}$ (Bower et al 2004).  A filamentary region of ionised gas,
Sgr A West of $\sim$ 2.5 pc in extent ($\sim$ 1$'$ ) surrounds Sgr A*
(Lo \& Claussen 1983; Roberts \& Goss 1993).
Both Sgr A* and Sgr A West are surrounded by a ring of molecular material
called the circumnuclear disk (CND) the outer edge of which has 
been traced up to $\sim$ 7 pc (Gusten et al 1987). Surrounding (in
projection) both Sgr A* and Sgr A West is the supernova remnant
Sgr A East with a radius of $\sim$ 5 pc (Ekers et al 1983, Maeda et al 2002). 
The supernova remnant is
in close proximity to Sgr A West and is expanding into
a molecular cloud complex (Pedlar et al 1989). 

The compact source Sgr A* and its surroundings have  been a target of HI 21 cm-line
measurements for many years. A number of absorption and emission 
features with a radial velocity range of
--190 km s$^{-1}$ to + 135 km s$^{-1}$ have been detected toward this region
which shows little Galactic rotation (Liszt et al 1983). However, many of the
components show evidence of non-circular motion.
Some of the anomalous absorption features are the 
--53 km s$^{-1}$ feature due to the $'$Expanding 3-kpc Arm$'$, the --135 km s$^{-1}$ 
feature due to the $'$Expanding Molecular Ring$'$, and the +50 km s$^{-1}$ feature due 
to the molecular cloud into which Sgr A East is expanding.

Early HI 21 cm-line observations toward  Sgr A* were carried out
using the Parkes Interferometer with a resolution of 3$^{'}$ (Radhakrishnan et al 1972). 
An analysis of the
HI 21 cm-line absorption spectrum toward Sgr A* revealed, apart from many familiar
features, an unexpected wide, shallow component (Radhakrishnan \& Sarma 1980, RS1). This 
component was centered at V$_{lsr}$ = --0.22 km s$^{-1}$ with a peak optical depth of 0.3 and
a velocity dispersion of 35 km s$^{-1}$ 
(full width at half maximum (FWHM) $\sim$ 80 km s$^{-1}$). 
This component was attributed to a new
population of shocked HI clouds in the Galaxy observed along the line of sight 
toward Sgr A* (Radhakrishnan \& Srinivasan 1980, RS2). Estimates indicated
that the kinetic energy in these clouds was $\sim$ 100 times that in the standard
HI clouds with consequent implications to the energetics of the 
interstellar medium. 

Subsequent HI absorption measurements toward Sgr A* using the 
Westerbork Synthesis Radio Telescope (WSRT) did not confirm the existence of this wide
line and placed an upper limit to its peak optical depth of 0.1(Schwarz, Ekers \& Goss 1982).
Around the same time, Shaver et al (1982), and Anantharamaiah et al 
(1984) made an analysis of the differences between the terminal velocities of HI
absorption spectra and the recombination line velocities
in the directions of 38 HII regions of known distances.
This analysis provided estimates of the number
densities and random velocity dispersions of interstellar HI clouds. Their results
supported the earlier postulate of shocked HI clouds in the Galaxy. 
However, Kulkarni \& Fich (1985) found that these results were easily confused
by many known systematic effects. After accounting for all these effects they found that
the results of Shaver et al (1982) and Anantharamaiah et al (1984) were quite uncertain.
Using HI emission data throughout the Galactic Plane,  
Kulkarni \& Fich (1985) suggested that the amount of HI in the high velocity
dispersion HI concentrations is an order of magnitude less than proposed
by RS2.

More recently, HI absorption measurements toward Sgr A* were carried out
using the Australia Telescope Compact Array (ATCA) 
(Rekhesh Mohan 2003, RM). These observations
confirmed the existence of a wide (FWHM $\sim$ 120 km s$^{-1}$) and shallow
($\tau_{peak} \sim$ 0.3) absorption feature toward Sgr A*. The discrepancy between
the earlier two observations can be attributed to the smaller velocity coverage
used in the WSRT observations.  The spatial distribution of this spectral feature
could not be obtained from the recent ATCA observations due to limited visibility coverage
and poor sensitivity.
A recent HI absorption measurement toward the Galactic Anticenter using the WSRT
detected no such wide line to a 3$\sigma$ optical depth limit of 0.006
indicating that the wide HI absorption line detected toward Sgr A* is not
ubiquitous (RM). Thus, the nature of the wide HI absorption line toward 
Sgr A* remained unclear.

From the earlier HI absorption measurements toward Sgr A* it is clear that
a large velocity coverage $\sim$ 600 km s$^{-1}$ is required to detect the wide 
HI absorption line. 
A velocity resolution $\sim$ 1 km s$^{-1}$ is also necessary to identify and remove the
narrow (FWHM $\sim$ a few km s$^{-1}$) HI absorption lines
which are detected along most lines of sight in the Galaxy and more so toward
Sgr A*.  In addition, 
an excellent visibility coverage is necessary to image the complex continuum and HI
distribution observed toward Sgr A* and its immediate surroundings. 
Therefore, we have undertaken a high spectral and spatial resolution
HI absorption study of the Sgr A* and its immediate surroundings ($\sim$ 5$'$)
 with the Very Large Array in order to determine the spatial
distribution of the shallow and wide HI absorption component, to characterize
its physical properties, and to determine its origin and physical relationship to
the center of the Galaxy.  In section 2, we present the observations, 
 in section 3 the data analysis, and in Section 4 the discussion.

\section{OBSERVATIONS \& IMAGING}

The observations were carried out using the Very Large Array (VLA) in the 36 km (A), 
11 km (B), 3 km (C), and 1 km (D) configurations during Jan - Oct, 2002. The total 
integration times in each of these configurations were $\sim$ 13, 9, 2 \& 2 hours respectively.
The field center of these observations was 
$\alpha$ (2000) = 17$^{h}$ 45$^{m}$ 40$^{s}$.049, and 
$\delta$ (2000) = --29$^{o}$ 00$^{'}$ 27$^{''}$.98.
The standard observing modes of the VLA spectrometer can not provide both the
required velocity coverage ($\sim$ 600 km s$^{-1}$) and velocity resolution
($\sim$ 1 km s$^{-1}$). Therefore, 
two overlapping bands each with 1.56 MHz bandwidth and 255 spectral
channels were arranged to cover the desired velocity range ($\pm$ 300 km s$^{-1}$)
 for the Galactic HI 21 cm-line.
There were 25 overlapping channels in the flat portions of the bands.
After precise bandpass calibration the two bands were combined together to produce 
a final data base with a velocity coverage of 590 km s$^{-1}$ at a spectral 
resolution of 1.3 km s$^{-1}$ after Hanning smoothing. 

\subsection{Imaging}

The calibration and imaging were carried out using the Astronomical Image Processing System
(AIPS). Using the data from the A+B+C+D arrays CLEANed image cubes containing
both the continuum and the spectral line data were produced.
Continuum images were produced from the respective image cubes by averaging the line-free
channels. Spectral line image cubes were produced by constructing a linear fit to the continuum 
in the line-free channels as a function of the channels and 
subtracting this linear fit from all channels.  Optical depth image cubes were
produced by suitably combining the spectral line image cubes and the respective continuum
images. A continuum image of the region surrounding Sgr A* is shown in Fig. 1. 
Representative optical depth
spectra at the four positions indicated in this figure are shown in Fig. 2.

Using the data from the C+D arrays, three sets of CLEANed image cubes containing
both the continuum and the spectral line data
were produced. These three data sets included all visibilities,  the visibilities 
beyond the inner 40 m, and the visibilities beyond the inner 100 m 
respectively. The upper limit to the included range of visibilities in all the three
images was $\sim$ 3400 m. Continuum images and optical depth image cubes were produced
following the procedure described earlier.
The contamination due to HI emission was minimal in all the images. 
However, of the three sets of images, the 
second set of images had the best spatial resolution and the maximum spatial extent
of the optical depth distribution. 
Since one of the primary motivation of the current observations
was to estimate the spatial distribution of the wide HI absorption line, the second
set of images was used in further analysis.
A continuum image of the region surrounding Sgr A* from the second data set
is shown in Fig. 3.
This image has a synthesized beam of 40 $'' \times$ 20$''$ (position angle = 0$^{o}$) with
an RMS of 30 mJy beam$^{-1}$.

The data from the A+B+C+D arrays provides higher spatial resolution compared to the
data from the C+D arrays. However, the radio continuum extent imaged from 
the high resolution data is rather limited (Fig. 1). The more extended continuum 
features detected in the C+D arrays data (Fig. 3) are resolved out in the high
resolution data. The limited extent of the radio continuum precludes the high
resolution images being used for tracing the spatial extent of the wide absorption
line. However, the high resolution data is used as an independent check of the
results from the low resolution data where both are available.

\section {ANALYSIS OF THE OPTICAL DEPTH SPECTRA AND IMAGES}

The motivation for the current observations are (a) to confirm the existence of a
wide velocity component in the HI absorption spectra toward Sgr A* and
(b) to understand its possible origin. The HI absorption
spectra in the Galactic plane are dominated by the presence of deep
(optical depth $\sim$ 0.5) and relatively small 
dispersion ($\sim$ a few km s$^{-1}$) lines. The presence of a large velocity 
dispersion ($\sim$ 30 - 50  km s$^{-1}$) HI absorption component
can manifest as a broad shoulder underlying the narrow lines in an optical 
depth spectrum. Often a visual inspection can reveal this shoulder. 
Representative optical depth spectra from the C+D array data at the positions
marked in Fig. 3 are displayed in Figs. 4 and 5. The spectra in Fig. 4 at 
positions within $\sim$ 2$'$ from Sgr A* reveal such a shoulder which is
absent in the spectra at positions outside $\sim$ 2 $'$ from Sgr A* (Fig. 5).
However, a quantitative 
analysis of the optical depth spectra is necessary to confirm these visual
impressions.

The HI 21 cm-line absorption or emission studies
carried out in the past indicate that most of these profiles are well fit by
Gaussians.  The Gaussian decomposition of the optical depth spectrum 
 was carried out by a code written for this purpose based on
the Levenberg-Marquardt Method (Numerical Recipes in Fortran 1992). 
This method minimizes the $\chi^{2}$ by computing
both its first and second derivatives with respect to the variables (parameters of the Gaussians
to be fitted) and approaching a minimum. 
The desired number of input Gaussians are fit to the data simultaneously and the best-fit
model is derived after a suitable number of iterations during which the parameters of the
input Gaussians are varied to obtain the minimum $\chi^{2}$. Two examples illustrating
such an analysis are given in Figs. 6 and 7. The data in Figs. 6 and 7 are the spectra marked
$'$1$'$ and $'$5$'$ in Figs. 4 and 5 respectively.
The best-fit model for
the data in Fig. 6 is displayed in the second panel from the top in Fig. 6. The third
panel from top is the residual spectrum (data-model), consistent with noise. The bottom panel
in Fig. 6 displays the residual when the 11 narrow lines in the model are
subtracted from the data. A wide line of FWHM $\sim$ 129 km s$^{-1}$ is 
detected in the data.  The 11 narrow components have a mean 
FWHM of 12$\pm$8 km s$^{-1}$, with the FWHMs in the range 4 -- 24 km s$^{-1}$. 
The best-fit Gaussian model for the data in Fig. 7 has 8 components with a 
mean FWHM of 7$\pm$3 km s$^{-1}$. No wide line is detected in this spectrum. The
quantitative analysis confirms the visual impressions of these spectra.

A Gaussian decomposition of the optical depth spectra from the C+D array data
 at independent positions across
the extent of the source (Fig. 3) was carried out. Such an analysis detected
a wide line (FWHM $\sim$ 120  km s$^{-1}$) in the spectra at positions 
within $\sim$ 2$'$ of Sgr A*.
No wide line was detected in the spectra at positions beyond $\sim$ 2$'$ of Sgr A*. 
Wide lines of similar characteristics were also detected in the spectra from the high
resolution images (Fig. 2). 

\subsection {Physical Features Associated with the Wide Line}

The position-velocity diagrams are an effective means to identify the physical
features associated with spectral lines. As an illustrative example,
 two declination-velocity 
images are shown in Figs. 8 and 9, corresponding to the positions marked
$'$1$'$ and $'$5$'$ respectively (Fig. 3). Both the figures are dominated by 
optical depth features which are parallel to declination and centered around
V$_{lsr} \sim$ 0, -25 and -50 km s$^{-1}$.  However, Fig. 8 shows
high optical depth features at positive velocities ($\sim$ 100 km s$^{-1}$) to the
north of Sgr A*, and at negative velocities ($\sim$ -100 km s$^{-1}$)
to the south of Sgr A*. These two features form two ends of an
inverted S-shaped feature (Fig. 8).  Detailed modeling has shown that these features 
result due to the circumnuclear
disk (CND) of mean radius 3.2 pc ($\sim$ 1.3$'$) rotating about Sgr A* with 
a velocity $\sim$ 100 km s$^{-1}$ (Liszt et al 1985). Such an inverted 
S-shaped feature is absent
in Fig. 9 due to the finite size of the CND.  The optical depth spectrum at the position
marked $'$1$'$ (Figs. 3 and 4) is the spectrum at $\delta$ = --29$^{o}$ 01$^{'}$ 03$^{''}$
in Fig. 8 and originates from the CND. The spectrum at the position 
marked $'$5$'$ (Figs. 3 and 5)
is the spectrum at $\delta$ = --28$^{o}$ 58$^{'}$ 50$^{''}$ in Fig. 9.

\section{DISCUSSION}

The current observations have demonstrated the existence of a wide line
($\delta$V$_{1/2} \sim$ 120 km s$^{-1}$) in the optical depth spectra 
toward Sgr A* and its surroundings, confirming the early
results of RS1. Since such a wide line is detected in the present observations
toward positions within $\sim$ 2$^{'}$ from Sgr A*, it is understandable that the 
Parkes Interferometer 
observations toward Sgr A* detected such a wide line even with a poorer resolution
of $\sim$ 3$'$. No wide line was detected in the current observations
at positions beyond $\sim$ 2$^{'}$ from Sgr A*.

A comparison of the optical depth spectra at positions 
within $\sim$ 2$^{'}$ from Sgr A* and the corresponding 
position-velocity diagrams indicates that the origin of the wide line 
is related to the CND. 
The wide line at the position marked $'$1$'$ (Fig. 6) results due to 
contributions from 
different parts of the CND in different velocity ranges (Fig. 8). 
On the other hand, the declination-velocity diagram 
corresponding to the position marked $'$5$'$ (Fig. 9) does not show the 
features corresponding 
to the CND as the CND is confined to within $\sim$ 2$^{'}$ of Sgr A*. Consequently,
the optical depth spectrum at the position marked $'$5$'$ (Fig. 7) does not
detect a wide line.

The wide line detected in the Parkes Interferometer observations toward Sgr A* 
was interpreted as a large velocity dispersion line (RS1). The dispersion
was attributed to the random motions of a new population of shocked HI clouds 
in the Galaxy observed along the line of 
sight toward Sgr A* (RS2). This new population was postulated to be
warmer and  more abundant ($\sim$ 15 clouds/kpc) than the standard cold HI
clouds. If such a scenario were correct, a diffuse feature over a range of velocities
corresponding to the FWHM of the wide line should have been detected in the 
declination-velocity diagrams (Figs. 8 and 9). 
No such feature was detected in these images.
The shocked HI cloud picture in this context is thus implausible.
Consequently, there are no implications of the wide line to the energetics of the
interstellar medium. The current observations have uncovered the origin
of the wide HI absorption line toward Sgr A* and have put to rest the 
related controversies that existed for over two decades.

\section*{acknowledgements}
The National Radio Astronomy Observatory (NRAO) is a facility of the National 
Science Foundation, operated under cooperative agreement by Associated
Universities, Inc..

\clearpage

\begin{figure*}
\epsfig{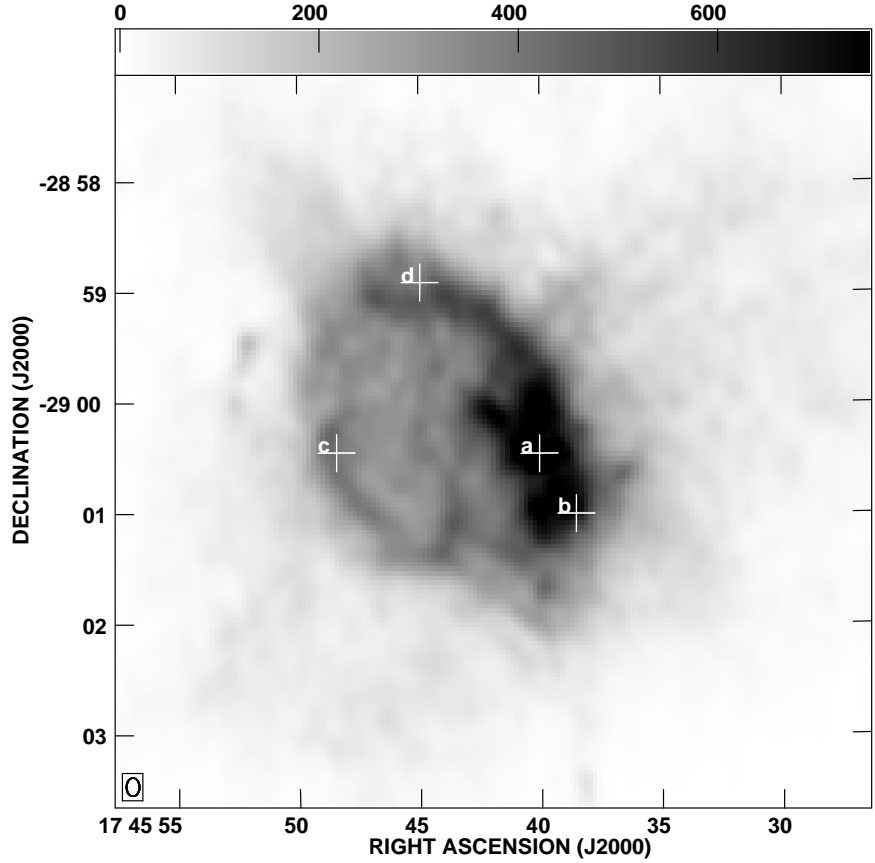}
\caption {A radio continuum image made from the A+B+C+D array data. The synthesised 
beam, 10$''$ X 7$''$ at a position angle of 0$^{o}$, is shown at the bottom
left hand corner. The RMS is 9 mJy/beam. The grey scale flux density range
is 0 to 750 mJy/beam.  The four crosses marked a, b, c, and d indicate the
positions at which the four spectra shown in Fig. 2 were obtained respectively.
The cross marked 'a' is the position of Sgr A*. }
\end{figure*}

\begin{figure*}
\epsfig{file=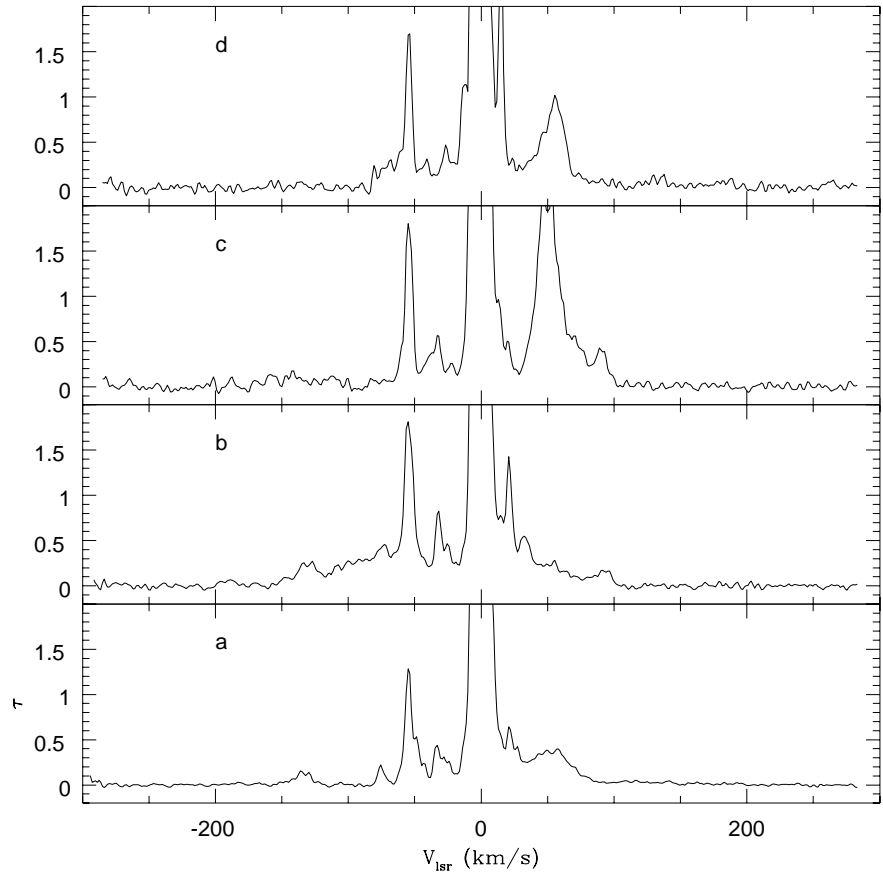, width=12cm}
\caption { Optical depth spectra from the A+B+C+D array data. The four spectra
correspond to the four positions marked in Fig. 1 respectively. 
Gaussian analysis of the spectra detects a wide line (FWHM $\sim$ 120 km s$^{-1}$)
apart from the many narrow lines that are evident.  In many 
spectra, like for e.g. the one marked 'b', such a wide line is evident as
a broad shoulder underneath the narrow lines.
}
\end{figure*}

\begin{figure*}
\epsfig{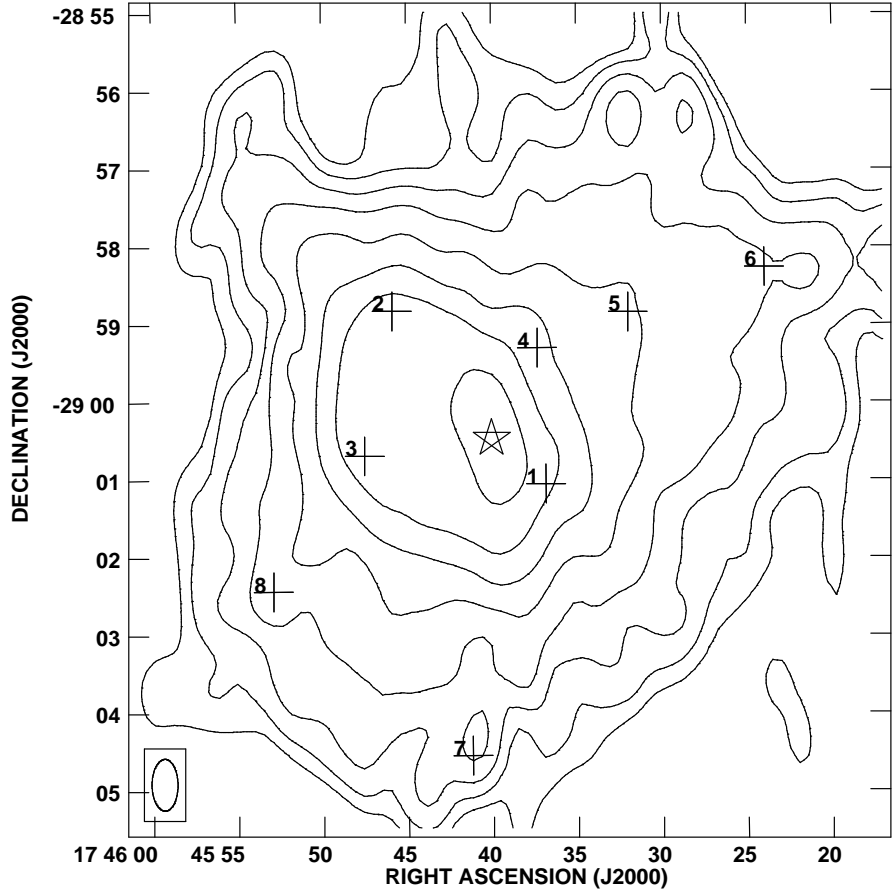}
\caption { A radio continuum image made from the C+D array data. The synthesised
beam, 40$''$ X 20$''$ at a position angle of 0$^{o}$, is shown at the bottom
left hand corner. The RMS is 30 mJy/beam. Contours start at 
50 mJy/beam and increase successively by a factor of 2. The star marks the
position of Sgr A*. The crosses mark the positions
at which the optical depth spectra shown in Figs. 4 and 5 were extracted. }
\end{figure*}

\begin{figure*}
\epsfig{file=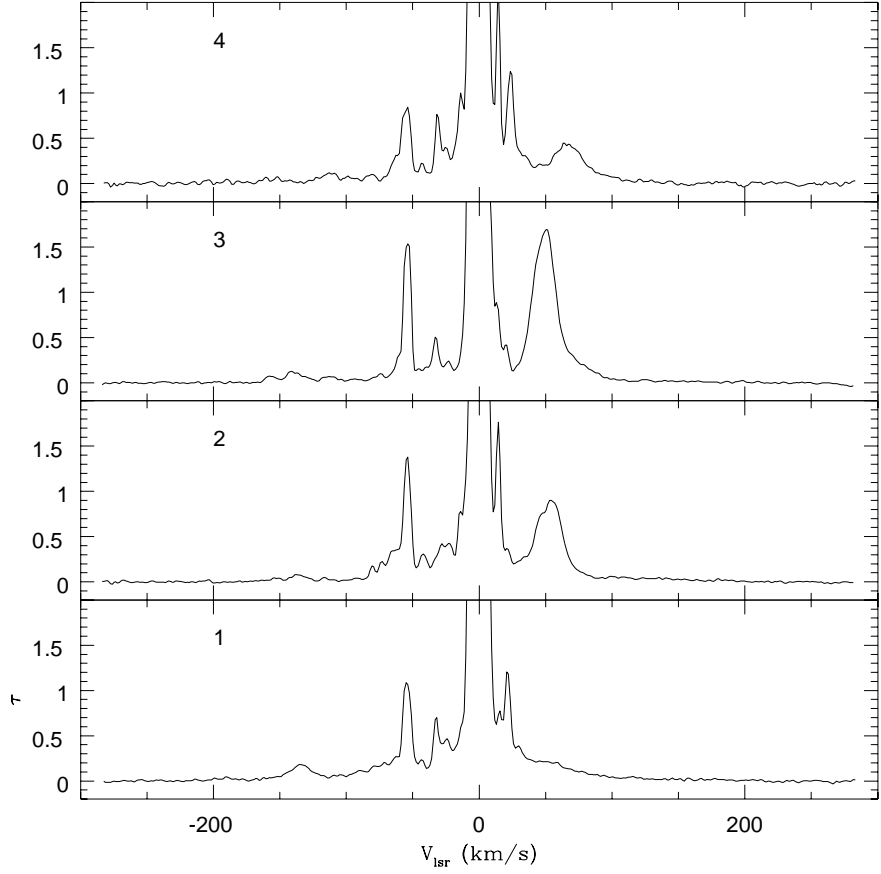, width=12cm}
\caption{Optical depth spectra from the C+D array data at the positions 1, 2, 3 and 4
respectively marked in Fig. 3.
These spectra are representative of the spectra at positions within $\sim$ 2 $'$ from Sgr A*.
A Gaussian analysis detects a wide line with mean values of V$_{lsr}$ = -4 $\pm$ 15
km s$^{-1}$, $\delta$V$_{1/2}$ = 119 $\pm$ 42 km s$^{-1}$, and $\tau_{peak}$ = 0.32
$\pm$ 0.12 in the spectra at positions within  $\sim$ 2 $'$ from Sgr A*. 
}
\end{figure*}

\begin{figure*}
\epsfig{file=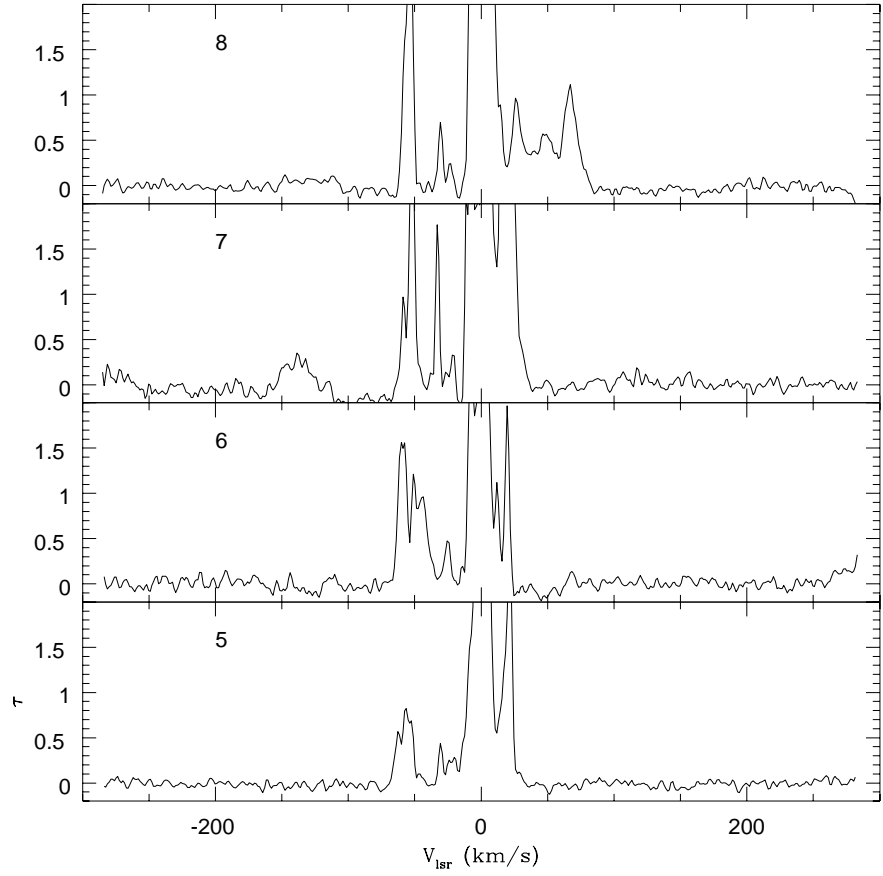, width=12cm}
\caption{Same as in Fig. 4 but for positions 5, 6, 7 and 8 respectively. 
These spectra are representative of the spectra at positions beyond $\sim$ 2 $'$ from Sgr A*.
A wide line of the kind detected in the spectra in Fig. 4 is absent in these spectra.}
\end{figure*}

\begin{figure*}
\epsfig{file=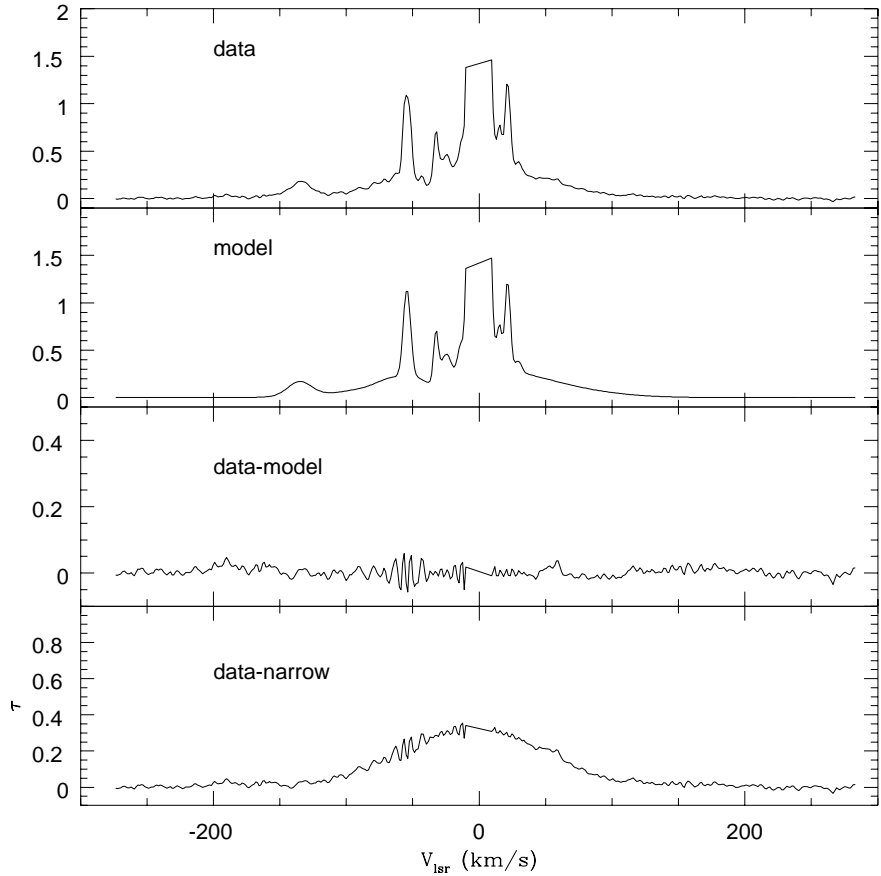, width=12cm}
\caption{Gaussian Decomposition. The data is the optical depth spectrum
from the C+D array data (marked '1' in Figs. 3 and 4). The best-fit model
consists of 11 narrow Gaussians with a mean FWHM of 12 $\pm$ 8 km s$^{-1}$
and a wide Gaussian of FWHM 129 $\pm$ 1 km s$^{-1}$. 
The residual (data-model) is consistent with noise. The bottom panel 
displays the residual when only the narrow lines in the model are subtracted
from the data. This wide line is also evident in the data as a broad
shoulder.}
\end{figure*}

\begin{figure*}
\epsfig{file=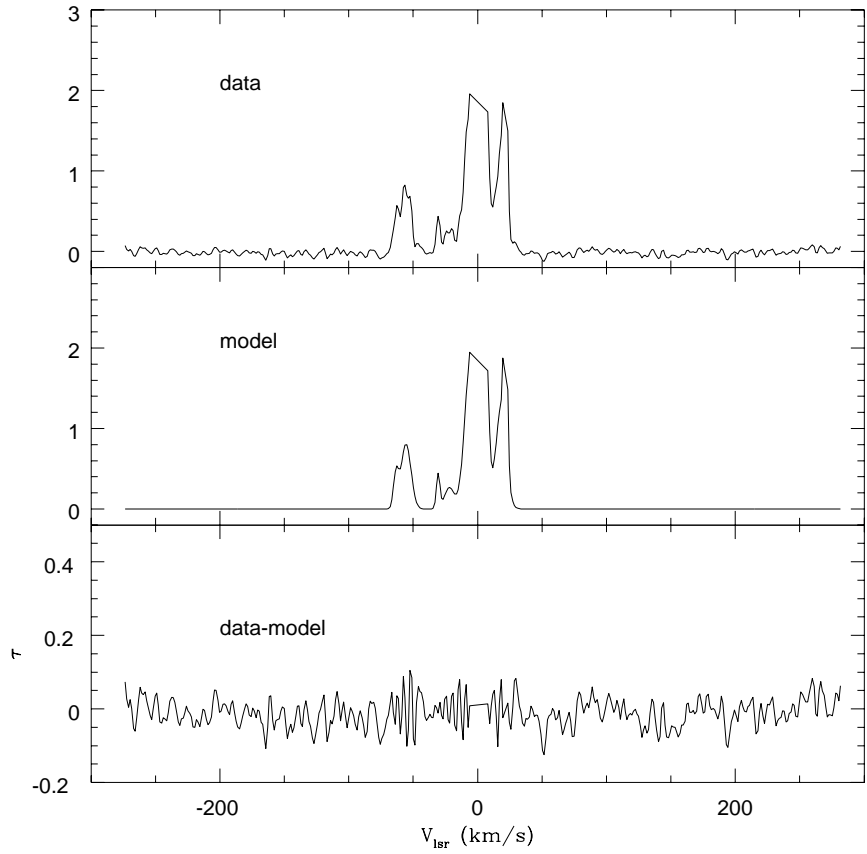, width=12cm}
\caption{Gaussian Decomposition. The data is the optical depth spectrum from the C+D array data
(marked '5' in Figs. 3 and 5). The best-fit model consists
of 8 narrow Gaussians with a mean FWHM of 7 $\pm$ 3 km s$^{-1}$.
The residual (data-model) is consistent with noise. No wide line is
detected in this spectrum.
}
\end{figure*}

\begin{figure*}
\epsfig{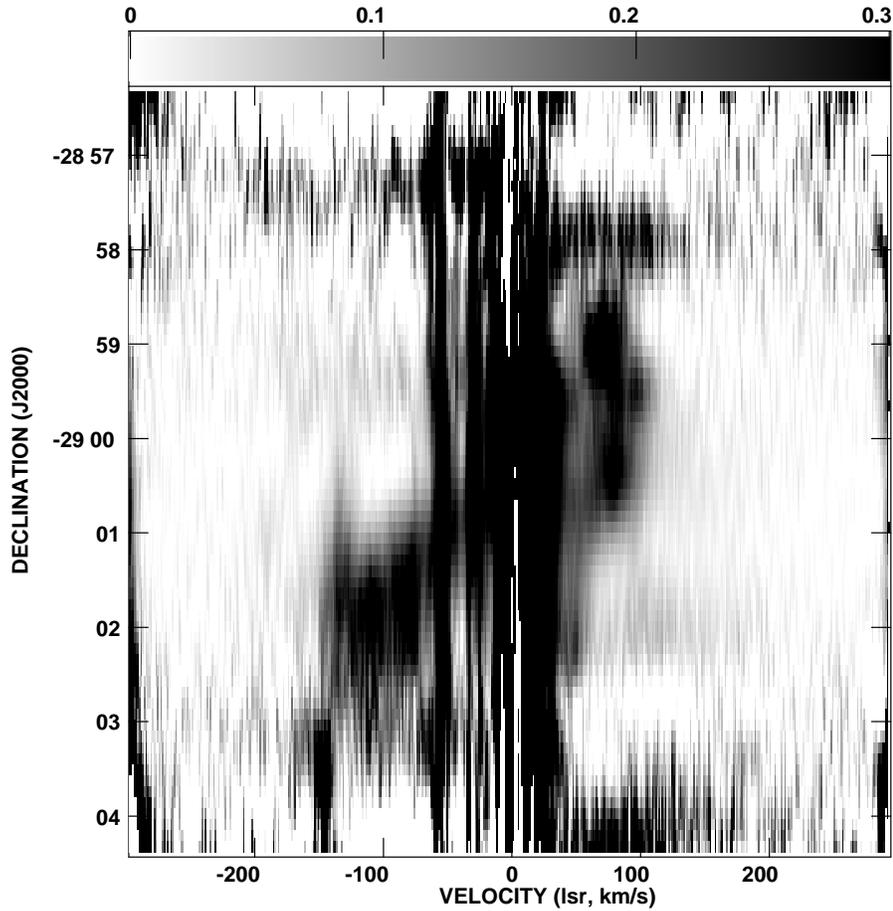}
\caption{Declination-Velocity diagram from the C+D array data
 at the right ascension of the position marked '1' in Fig. 3.
The grey scale is optical depth in the range 0 - 0.3. The reliable
portion of the image is --29$^{o}$ 3$^{'} < \delta <$ --28$^{o}$ 57$^{'}$. The 
features to the north of Sgr A* at positive velocities ($\sim$ 100 km s$^{-1}$)
 and to the south of Sgr A*
at negative velocities ($\sim$ --100 km s$^{-1}$) are due to the circumnuclear 
disk surrounding Sgr A*.
}
\end{figure*}

\begin{figure*}
\epsfig{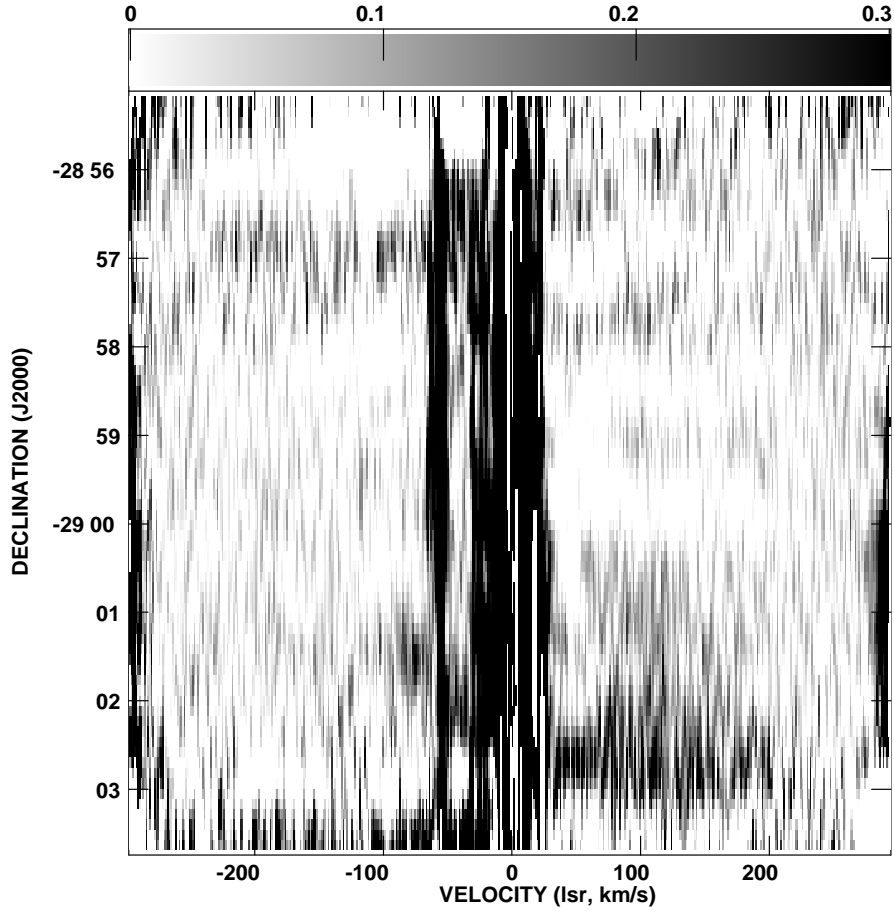}
\caption{ Same as in Fig. 8 but at the right ascension of the position marked
'5' in Fig. 3.
The CND is not detected in this image since the CND
is confined to within $\sim$ 2$'$ from Sgr A*.
}
\end{figure*}

\clearpage


\clearpage

\begin{thebibliography}{10}

\bibitem{} Anantharamaiah, K. R., Radhakrishnan, V., \& Shaver, P. A. 1984, AA, 138, 131.

\bibitem{} Bower, G.C., Falcke, H., Herrnstein, R.M., Zhao, J.H., Goss, W.M., 
Backer, D.C. 2004, Science, 304, 704.



\bibitem{} Ekers, R. D., van Gorkom, J. H., Schwarz, U. J., \& Goss, W. M.
1983, AA, 122, 143.



\bibitem{} Gusten, R., Genzel, R., Wright, M. C. H., Jaffe, D. T., Stutzki, J., Harris, A. I.
1987, ApJ, 318, 124.


\bibitem{} Kulkarni, S. R., \& Fich, M. 1985, ApJ, 289, 792.

\bibitem{} Liszt, H. S., van der Hulst, J. M., Burton, W. B., \& Ondrechen, M. P.
1983, AA, 126, 341.

\bibitem{} Liszt, H. S., Burton, W. B., van der Hulst, J. M. 1985, AA, 142, 237.

\bibitem{} Lo, K. Y., \& Claussen, M. J. 1983, Nature, 306, 647.
 
\bibitem{} Maeda, Y et al 2002, ApJ, 570, 671.



\bibitem{} Numerical Recipes in FORTRAN, Press, W. H., Teukolsky, S. A.,
Vetterling, W. T., \& Flannery, B. P., Cambridge University, 1992.

\bibitem{} Pedlar, A., Anantharamaiah, K. R., Ekers, R. D., Goss, W. M., 
\& van Gorkom, J. H. et al 1989, ApJ, 342, 769.

\bibitem{} Radhakrishnan, V., Goss, W. M., Murray, J. D., \& Brooks, J. W. 1972 ApJS, 24, 49.

\bibitem{} Radhakrishnan, V., \& Sarma, N. V. G. 1980, AA, 85, 249 (RS1).

\bibitem{} Radhakrishnan, V., \& Srinivasan, G. 1980, JAA, 1, 47 (RS2).

\bibitem{} Rekhesh Mohan 2003, PhD Thesis, Jawaharlal Nehru University, New Delhi (RM).

\bibitem{} Roberts, D. A., \& Goss, W. M. 1993, ApJS, 86, 133.

\bibitem{} Schwarz, U. J., Ekers, R. D., \& Goss, W. M. 1982, AA, 110, 100.

\bibitem{} Shaver, P. A., Radhakrishnan, V., Anantharamaiah, K. R., 
Retallack, D. S., Wamsteker, W., \& Danks, A. C. 1982, AA, 106, 105.

\end{thebibliography}
\end{document}